\documentclass[preprint,english,showpacs,12pt,floatfix]{revtex4-1}
\usepackage{babel,amsmath,amssymb,dcolumn}
\usepackage{hyperref}
\usepackage{amsfonts}
\usepackage{amssymb}
\usepackage{graphicx}
\usepackage{latexsym}
\usepackage{dcolumn}
\usepackage{epsfig}
\usepackage{subfigure}
\begin{document}

\newcommand{\vp}{\varphi}
\newcommand{\nn}{\nonumber\\}
\newcommand{\beq}{\begin{equation}}
\newcommand{\eeq}{\end{equation}}
\newcommand{\bed}{\begin{displaymath}}
\newcommand{\eed}{\end{displaymath}}
\def\bea{\begin{eqnarray}}
\def\eea{\end{eqnarray}}
\newcommand{\veps}{\varepsilon}
\newcommand{\nablasl}{{\slash \negthinspace \negthinspace \negthinspace \negthinspace  \nabla}}
\newcommand{\om}{\omega}

\newcommand{\Dsl}{{\slash \negthinspace \negthinspace \negthinspace \negthinspace  D}}
\newcommand{\tDsl}{{\tilde \Dsl}}
\newcommand{\tnablasl}{{\tilde \nablasl}}
\title{Fermion perturbations in string-theory black holes}

\author{Owen Pavel Fern\'{a}ndez Piedra}
%%%%%%%%%%%%%%%%%%%%%%%%%%%%%%%%%%%%%%%%%%%%%%%
\email{opavel@ucf.edu.cu }
%%%%%%%%%%%%%%%%%%%%%%%%%%%%%%%%%%%%%%%%%%%%%%%
\affiliation{$^{1}$ Instituto de F\'{\i}sica, Universidade de S\~ao Paulo,
  CP 66318,
05315-970, S\~ao Paulo, Brazil ,}
\affiliation{$^{2}$ Departamento de F\'{i}sica y Qu\'{i}mica, Universidad
de Cienfuegos, Carretera a Rodas, Cuatro Caminos, s/n. Cienfuegos,
Cuba}

\author{Jeferson de Oliveira}
\email{jeferson@fma.if.usp.br }
\affiliation{Instituto de F\'{\i}sica, Universidade de S\~ao Paulo,
  CP 66318,
05315-970, S\~ao Paulo, Brazil}

%%%%%%%%%%%%%%%%%%%%%%%%%%%%%%%%%%%%%%%%%%%%%%%
%%%%05315-970, S\~ao Paulo, Brazil}

\begin{abstract}
In this paper we study fermion perturbations in four dimensional black holes
of string theory, obtained either from a non-extreme configuration of three intersecting five-branes 
with a boost along the common string or from a non-extreme intersecting system of two two-branes and two five-branes. 
The Dirac equation for the massless neutrino field, after conformal re-scaling of the metric, is written as a wave equation 
suitable to study the time evolution of the perturbation. With the aid of Prony fitting of time-domain profile we calculate 
the complex frequencies that dominate the quasinormal ringing stage, and also determine this quantities by the semi-analytical
sixth order WKB method. We also find numerically the decay factor of fermion fields at very late times, and show that the falloff 
is identical to those appeared for massless fields in other four dimensional black hole spacetimes. 
\end{abstract}

\preprint{GFTeor-UCF 2010-04}
\date{\today}
\maketitle

%%%%%%%%%%%%%%%%%%%%%%%%%%%%%%%%%%%%%%
\section{Introduction}
String theory and its generalization in terms of extended objects named  M/Dp-branes is a promising candidate for a fundamental quantum theory of all interactions \cite{pol}, including a description of black holes in a quantum gravity framework \cite{Maldacena:1996ky}. 

The scenario of string theory is consistently formulated only  in higher dimensions, which are usually compactified, so a full description of a black hole whose event horizon is comparable to the size of extra dimensions has to be in terms of such higher dimensional description.  So, in order to discuss the production of small black holes in the Large Hadron Collider (LHC) and the evaporation of black holes by Hawking effect, we must consider solutions coming from a string theory/branes picture.

One class of solutions which encompass the features of string theory and also can be interpreted as black holes are those non-extremal solutions which came from intersections of branes in string theory or in M-theory \cite{cvetic1}\cite{cvetic2}\cite{cvetic3}\cite{horowitz1}.  The parameters of such black holes can be deduced by corresponding compactifications of the higher dimensions.

In this work we aim at studying the question of stability of a family of  four-dimensional black holes in string theory.  In order to perform this we calculate the quasinormal spectrum due to a perturbation of a Dirac field evolving in the background given by the stringy black hole. Regarding the stability of extended objects from low energy limit of string theory, some work has been done \cite{nuestro}, where it was considered a probe scalar field evolving in a $p$-brane geometry and its quasinormal spectrum in order to analyze the stability and the relation between the parameters of $p$-brane and the quasinormal ringing.

Quasinormal modes (QNM) of black holes has been extensively studied since the pioneering work on stability of Schwarzschild singularity performed by Regge and Wheeler \cite{Regge:1957td}. In studying quasinormal spectrum, we can gain some valuable information about the parameters which caracterize the solution, because there are definite relations between the QNM and the parameters of solution, see \cite{Berti:2009kk} and references therein. Such QNM are characterized  by a well-defined set of complex frequencies and  encode the response of  black hole to external perturbation, so we can study the stability of a black hole against small perturbations due to probe fields or the geometry itself trough its quasinormal spectrum.

Besides, further developments in QNM research in the context of AdS/CFT and holography allow us to calculate the location of the poles of the retarded correlators functions of certain gauge theories \cite{Son:2007vk} and reveals some connection between the dynamics of  black hole horizons and the hydrodynamics \cite{Gubser:2009md} \cite{Hartnoll:2009sz}.

The paper is organized as follows: Section II gives  a review about the casual structure of four dimensional black hole obtained from intersections of branes.
 In Section III we obtain the wave equation suitable to analyze the propagation of a  Dirac field in the background geometry of the four dimensional stringy black hole. Section IV is devoted to numerically solve the evolution
equation of such fields in the considered background, and in Section V we will concerned with the quasinormal stage, and compute the complex quasinormal frequencies using two methods: Prony fitting of
time domain data and the sixth order WKB approach. The results of a numerical investigation of the relaxation of Dirac perturbations in stringy black holes at very long times are presented in Section VI. In Section VII we present an analytical expression for the quasinormal frequencies in the limit of large angular momentum, followed by the last section of the paper, which contains some concluding remarks of our work.

%%%%%%%%%%%%%%%%%%%%%%%%%%%%%%%%%%%%%
\section{(3+1)-dimensional black hole solution from intersecting branes}
The metric of  the non-extremal 4-dimensional black hole obtained from intersections of five-branes read as \cite{cvetic2}
\begin{equation}\label{metric}
ds^{2}=f^{-1/2}\left(1-\frac{r_{H}}{r}\right)dt^{2}+f^{1/2}\left[\left(1-\frac{r_{H}}{r}\right)^{-1}dr^{2}+r^{2}d\Omega_{2}^{2}\right],
\end{equation}
where
\begin{equation}\label{f}
f=(1+\frac{r_{H}Q_{1}}{r})(1+\frac{r_{H}Q_{2}}{r})(1+\frac{r_{H}Q_{3}}{r})(1+\frac{r_{H}Q_{4}}{r}).
\end{equation}

This metric represents a family of black hole solutions parametrized by $r_{H}$, which gives the location of the event horizon, and  four charges $Q_{1}$, $Q_{2}$, $Q_{3}$, $Q_{4}$  given by :
\begin{eqnarray}\label{charges}\nonumber
Q_{i}=\sinh^{2}(\delta_{i}),\hspace{0.3cm} i=1,2,3,4
\end{eqnarray}
that are written in terms of five-brane parameters $\delta_{i}$, resulting from the compactification of higher dimensions. This solution describes a four dimensional Schwarzschild black hole in the case which the charges $Q_{i}$ are all zero. 
Also the metric is asymptotically  flat and, as we shall see below, has a regular event horizon located at $r=r_{H}$ even with all charges different from zero. Besides, the redshift goes to infinity in such surface. 

Infinity redshift surfaces can be found, for a given metric, trough the following relation
\begin{equation}\label{redshift}
\nu=\nu_{0}\sqrt{\frac{g_{00}(x^{a}_{source})}{g_{00}(x^{a})}},
\end{equation}
which relates the frequency $\nu$ measured by an observer at rest away from the source, whose emission frequency of say, light pulses,  is $\nu_{0}$. 

In order to have a infinity redshift surface, the frequency $\nu$ has to be zero, it means the frequency $\nu$ was infinitely delayed due to gravitational effects. So,  equation(\ref{redshift}) implies
\begin{equation}\label{redshift2}
g_{00}(x^{a}_{source})=0,
\end{equation}
given us the location of the infinity redshift surfaces. In the case of the spacetime of four dimensional stringy black holes (\ref{metric}) it is easy to see that $r=r_{H}$ and $r=0$ are the surfaces where the 
redshift of in-falling objects goes to infinity.

As in Schwarzschild solution, such surface $r=r_{H}$ acts as a one-way membrane for physical objects whose trajectories lie in or on the forward light cone. To see this, suppose a smooth hypersurface $S$ defined by the equation

\[
u(x^{a})=const.
\]
The vector $N_{a}=\partial_{a}u$ is normal to $S$, so if $S$ is a one-way membrane it has to be a null hypersurface, therefore the norm of  $N_{a}$ has to be null as well.  So,
\[
g^{ab}N_{a}N_{b}=0,
\]
determine the one-way membranes, which in our case are located at $r=r_{H}$ and $r=0$. However, for the one of the  four charges $Q_{i}$ equal to zero, only $r=r_{H}$ is a null and infinity redshift hypersurface, $r=0$ is a genuine space-time singularity, as we will see below in more detail.

Let us consider the Kretschmann invariant for the spacetime (\ref{metric}), where we take for simplicity $Q_{i}=q$ (the result holds  as well for four different charges):
\begin{equation}\label{inv}
R^{abcd}R_{abcd}=\frac{4}{(r_{H}q+r)^{8}}P(r),
\end{equation}
where 
\[
P(r)=r_{H}^{4}q^{2}(5+4q+q^{2})-2r_{H}^{3}q r(3+7q+2q^{2})+(r_{H}r)^{2}(3+8q+8q^{2})-2r_{H}r^{3}+r^{4}.
\]
As we see from the above expressions, for $r\rightarrow r_{H}$ we have

\[
R^{abcd}R_{abcd}(r\rightarrow r_{H})\rightarrow const,
\]

and for $r\rightarrow 0$
 
 \begin{equation}\label{inv_0}
R^{abcd}R_{abcd}(r\rightarrow 0)\rightarrow \frac{4}{r_{H}^{8}q^{8}}\left[r_ {H}^{4}q^{2}(5+4q+q^{2})\right].
\end{equation}

We see clearly that $r=0$ is not a spacetime singularity but, as previously pointed out by Horowitz {\it{et al}} \cite{horowitz} it is  actually a inner horizon.
 However, if at least one of the charges is zero, the surface $r=0$ becomes singular as we see explicitly in the expression (\ref{inv_0}).

We conclude that for at least one charge $Q_{i}$  zero, we have a four dimensional black hole with a spacetime singularity located at $r=0$ covered by a
 event horizon at $r=r_{H}$ which is also a infinity redshift surface, as expected for a spherically symmetric  space time. On the other hand, if all charges 
are non-zero, we have a regular black hole with a event horizon at $r=r_{H}$ and a inner horizon at $r=0$.

%%%%%%%%%%%%%%%%%%%%%%%%%%%%%%%%%%%%%
\section{Fundamental equations}

In curved space-time the massless Dirac equation is written as
\begin{equation}\label{}
   \nablasl \Psi=0,
\end{equation}
where $\nablasl=\Gamma^{\mu}\nabla_{\mu}$ is the Dirac operator that acts on the four-spinor $\Psi$, $\Gamma_{\mu}$ are the curved space Gamma matrices, and the covariant derivative is 
defined as $\nabla_{\mu}=\partial_{\mu}-\frac{1}{4}\om_{\mu}^{a b}\gamma_{a}\gamma_{b}$, with $\mu$ and $a$ being tangent and space-time indices respectively, related by the basis of orthonormal one forms 
$\vec{e}^{\ a}\equiv e_{\mu}^{a}$. The associated conection one-forms $\om_{\mu}^{a b}\equiv\om^{a b}$ obey $d\vec{e}^{\ a}+\om^{a}_{\ b}\wedge\vec{e}^{\ b}=0$, and the $\gamma^{a}$ are flat space-time gamma matrices 
related with curved-space ones by $\Gamma^{\mu}=e^{\mu}_{\ a}\gamma^{a}$. They form a Clifford algebra
in d dimensions, i e, they satisfy the anti-conmutation relations $\{\gamma^{a},\gamma^{b}\}=-2 \eta^{ab}$, with $\eta^{00}=-1$. 

\par First consider some properties of the Dirac operator that will be used in future calculations \cite{Gibbons,gibbons-rogatko}. Under a conformal transformation of the metric of the form :
\begin{eqnarray}
 & g_{\mu\nu}=\Omega^{2}\tilde{g}_{\mu\nu} , 
\end{eqnarray}
the spinor $\psi$ and the Dirac operator transforms as
\begin{eqnarray}
\psi =\Omega^{-\frac{3}{2}}\tilde{\psi} ,\\
\nablasl \psi =\Omega^{-\frac{5}{2}} \ \tilde{\nablasl}\tilde{\psi} .
\end{eqnarray}
If the line element of a space-time metric takes the form of a sum of independent components $ds^{2}=ds_{1}^{2}+ds_{2}^{2}$, where $ds_{1}^{2} =g_{ab} (x) dx^a dx ^b$ and $ds_{2}^{2}= g_{mn}(y) dy^m dy ^n$
then Dirac operator $\nablasl$ satisfies a  direct sum decomposition
\beq
\nablasl = \nablasl_x + \nablasl_y.
\eeq
The above mentioned properties of spinors and the Dirac operator in direct sum metrics and conformal related ones, allows a simple treatment of the massless Dirac equations in curved spacetimes. For two 
conformally related metrics, the validity of massless Dirac equation in one
implies the the validity of the same equation in the other. Then, the idea is to solve the Dirac equation in the curved space described by an initial metric tensor performing successive conformal transformations
that isolate the metric components that depend of a given variable, and applied successively the direct sum decomposition of Dirac operator, until to obtain an equivalent problem in a spacetime of the form $M^{2}\times\sum^{2}$
, where $M^{2}$ is two-dimensional Minkowski spacetime and $\sum^{2}$ is a two-dimensional maximally symmetric one, in which the spectrum of massless Dirac operator is known. 
 
The method work as follows. First, in the following we suppose that our four dimensional metric is spherically symmetric, and given by:
\begin{equation}
ds^{2}=g_{\mu \nu}dx^{\mu}dx^{\nu}=-A(r)dt^{2}+B(r)dr^{2}+C(r)d\Omega^{2}_{2} ,
\end{equation}
where $d\Omega^{2}_{2}$ denotes the metric for the $(2)$-sphere $S^{2}$.

It is easy to show that with the identification $(t,r,\theta,\phi)\rightarrow{(0,1,2,3)}$ if we choose the basis one-forms as $\vec{e}^{\ 0}=\vec{e}^{\ 0}(t,r)=f_{(t)}(t,r)dt+f_{(r)}(t,r)dr$, $\vec{e}^{\ 1}=\vec{e}^{\ 1}(t,r)=g_{(t)}(t,r)dt+g_{(r)}(t,r)dr$ and 
$\vec{e}^{\ k }=\vec{e}^{\ k}(\theta,\phi)=h^{(k)}_{(\theta)}(\theta,\phi)d\theta+h^{(k)}_{(\phi)}(\theta,\phi)d\phi$, $k=2,3$ then we have some conection one-forms that are equal to zero, and only $\om^{0}_{\ 0}$, $\om^{0}_{\ 1}$, $\om^{1}_{\ 0}$,
 $\om^{1}_{\ 1}$ and $\om^{i}_{\ j}; i,j=2,3$ are different from zero. This fact allows us to write $\nabla_{a}=\nabla_{a}\textbf{I}_{(2)}; a=0..3$, where $\textbf{I}_{(2)}$ is the unit matrix in two dimensions \cite{lopez-ortega}.

Now, under a conformal reescaling of the form $ds^{2}=C(r)d\tilde{s}^{2}$, where:

\beq
d\tilde{s}^{2} = -\frac{A}{C}dt^{2} + \frac{B}{C}dr^{2} + d\Omega^{2}_{2} ,
\label{metric2}
\eeq
we need to solve the equation $\tilde{\nablasl}\tilde{\psi}=\tilde{\Gamma}^{\mu}\tilde{\nabla}_{\mu}\tilde{\psi}=0$, with $\tilde{\psi}=C^{\frac{3}{2}}\psi $. In general, the Dirac matrices $\tilde{\Gamma}^{\mu}$ can be choosen as:
\begin{eqnarray}
\tilde{\Gamma}^{0}=\tilde{\Gamma}^{0}_{(2)}\otimes\textbf{I}_{(2)}, \\
\tilde{\Gamma}^{1}=\tilde{\Gamma}^{1}_{(2)}\otimes\textbf{I}_{(2)}, \\
\tilde{\Gamma}^{2}=\tilde{\Gamma}^{5}_{(2)}\otimes\tilde{\Gamma}^{0}_{(2)}, \\
\tilde{\Gamma}^{3}=\tilde{\Gamma}^{5}_{(2)}\otimes\tilde{\Gamma}^{1}_{(2)}, 
\end{eqnarray}
where $\tilde{\Gamma}^{0}_{(2)}$, $\tilde{\Gamma}^{1}_{(2)}$ and $\tilde{\Gamma}^{5}_{(2)}=-\tilde{\Gamma}^{0}_{(2)}\tilde{\Gamma}^{1}_{(2)}$ are two-dimensional Gamma matrices, and $(\tilde{\Gamma}^{5}_{(2)})^{2}=\textbf{I}_{(2)}$. 

Since the orbit-space part and the angular part of the metric are completely separated, one can write the Dirac equation in the form:
\begin{eqnarray}
\left[ \left(
\tilde{\Gamma}^{0}_{(2)}\tilde{\nabla}_{0} +
\tilde{\Gamma}^{1}_{(2)}\tilde{\nabla}_{1} \right) \otimes \textbf{I}_{(2)}
 + \tilde{\Gamma}^{5}_{(2)} \otimes
\left( \tilde{\Gamma}^{a}
\tilde{\nabla}_{a}\right)_{S_{2}} \right] \tilde{\psi} = 0
, &
\end{eqnarray}
where $\left( \tilde{\Gamma}^{a}\tilde{\nabla}_{a}\right)_{S_{2}}$ denotes de Dirac operator in the $2$-sphere, whose orthogonal set of eigenspinors $\tilde{\xi}_{\ell}^{(\pm)}$ are defined by \cite{Camporesi}:
\begin{equation}
\left( \tilde{\Gamma}^{a}\tilde{\nabla}_{a} \right)_{S_{2}}\tilde{\xi}_{\ell}^{(\pm)} = \pm i \left(\ell  + 1\right) \tilde{\xi}_{\ell}^{(\pm)} ,
\end{equation}
where $l = 0, 1, 2, \dots$. Now expanding $\tilde{\psi}$ as:
\begin{equation}
\tilde{\psi} = \sum_{\ell} \left(\tilde{ \varphi}_{\ell}^{(+)} \tilde{\xi}_{\ell}^{(+)} +\tilde{ \varphi}_{\ell}^{(-)}\tilde{ \xi}_{\ell}^{(-)} \right) .
\end{equation}
we can put the Dirac equation in the form:
\begin{equation}\label{eqn:2Ddirac}
\left \{ \tilde{\Gamma}_{(2)}^{0}\tilde{ \nabla}_{0} + \tilde{\Gamma}_{(2)}^{1}\tilde{ \nabla}_{1} + \tilde{\Gamma}_{(2)}^{5} \left[ \pm i \left( l + 1 \right) \right] \right\} \tilde{\varphi}_{\ell}^{(\pm)} = 0 ,
\end{equation}

In the following we shall work with the $+$ sign solution, because the $-$ sign case can be worked in the same form. Denoting as $\tilde{ds_{1}}^{\ 2}=-\frac{A}{C}dt^{2} + \frac{B}{C}dr^{2}$ the $t-r$ part of
the metric (\ref{metric2}) and doing a conformal rescaling of it in the form:

\beq
d\tilde{s}_{1}^{\ 2}=\frac{A}{C} \left[ -dt^{2} + dr_{*}^{2}\right],
\eeq 
where $dr_{*}=\sqrt{\frac{B}{A}}dr$ is the tortoise coordinate, we have:
\begin{equation}\label{eqn:2Ddirac}
\left [ \gamma^{t}\partial_{t} +  \gamma^{r_{*}}\partial_{r_{*}} + i \gamma^{5} \sqrt{\frac{A}{C}}\left( \ell + 1 \right) \right] \tilde{\varphi}_{\ell}^{(+)} = 0 ,
\end{equation}
being $\gamma^{t}$ and $\gamma^{r_{*}}$ Dirac matrices in the two-dimensional Minkowski space-time with metric $ds^{2}=-dt^{2}+dr_{*}^{2}$. 

Now choosing the representation of Dirac matrices given by:
\begin{equation}
\gamma^{t} = -i\sigma^{3}\ \ \ ,\ \ \
\gamma^{r_{*}}= \sigma^{2} \ \ \ ,\ \ \
\gamma^{5} = (-i\sigma^{3})(\sigma^{2}) = - \sigma^{1},
\end{equation}
where the $\sigma^{i}$ are the Pauli matrices:
\begin{equation}
\sigma^{1}=\left(
\begin{array}{cc}
0 & 1 \\ 1 & 0
\end{array}
\right)\ \ \ ,\ \ \ \sigma^{2}=\left(
\begin{array}{cc}
0 & -i \\ i & 0
\end{array}
\right)\ \ \ ,\ \ \ \sigma^{3}=\left(
\begin{array}{cc}
1 & 0 \\ 0 & -1
\end{array}
\right), 
\end{equation}
and writting 
\begin{equation}
\tilde{\varphi}_{\ell}^{(+)} = \left(
\begin{array}{c}
i\zeta_{\ell}(t,r) \\ \chi_{\ell}(t,r)
\end{array}
\right) ,
\end{equation}
we obtain the following equations for each component of the Dirac spinor $\tilde{\varphi_{\ell}}$:
\begin{equation}
i\frac{\partial\zeta_{\ell}}{\partial t}+\frac{\partial\chi_{\ell}}{\partial r_{*}}+\Lambda_{\ell}\chi_{\ell}=0,
\end{equation}
and 
\begin{equation}
i\frac{\partial\chi_{\ell}}{\partial t}-\frac{\partial\zeta_{\ell}}{\partial r_{*}}+\Lambda_{\ell}\zeta_{\ell}=0,
\end{equation}
where 
\beq \label{lambda}
\Lambda_{\ell}(r)= \sqrt{\frac{A}{C}}\left( \ell + 1 \right) .
\eeq
The above equations can be separated to obtain:
\begin{equation}
\frac{\partial^{2}\zeta_{\ell}}{\partial t^{2}}-\frac{\partial^{2}\zeta_{\ell}}{\partial r_{*}^{2}}+ V_{+}(r)\zeta_{\ell}=0,
\label{tevol1}
\end{equation}
and 
\begin{equation}
\frac{\partial^{2}\chi_{\ell}}{\partial t^{2}}-\frac{\partial^{2}\chi_{\ell}}{\partial r_{*}^{2}}+ V_{-}(r)\chi_{\ell}=0,
%\label{tevol2}
\end{equation}
where:
\begin{equation}
V_{\pm}=\pm\frac{d\Lambda_{\ell}}{dr_{*}}+\Lambda_{\ell}^{2} .
\label{pot}
\end{equation}
The above equations gives the temporal evolution of Dirac perturbations outside the black hole spacetime \cite{splitfermion}. As the potentials $V_{+}$ and $V_{-}$ are supersymmetric to each other in the sense considered by Chandrasekhar in \cite{chandra}, then $\zeta_{\ell}(t,r)$ and $\chi_{\ell}(t,r)$ will have similar time evolutions and 
then they will have the same spectra, both for scattering and quasi-normal. At this point it should be stressed that  for the spinor $\tilde{\varphi}_{\ell}^{(-)}$, we have these two potentials again. In the following we will work with equation (\ref{tevol1}) and we 
eliminate the subscript $(+)$ for the effective potential, defining $V(r)\equiv V_{+}(r)$ .
%%%%%%%%
\begin{figure}[t]
\begin{center}
%\scalebox{0.95}{\includegraphics{potenciales_dirac_stringy.eps}}
\scalebox{0.95}{\includegraphics{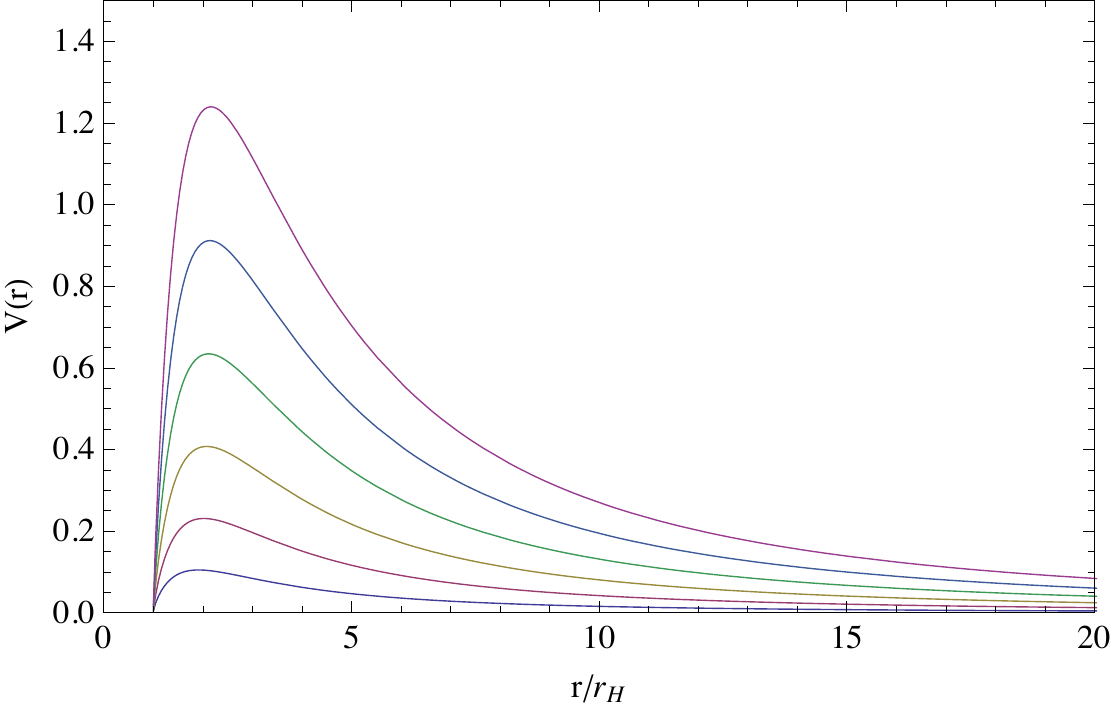}}
\end{center}
\caption{Potential $V(r)$ for Dirac perturbations with $\ell=0$ (bottom) to $\ell=5$(top).}
\label{potencialesdirac}
\end{figure}
%%%%%%%%%

Now for (3+1)-string theory black holes with line element (\ref{metric}), making the identifications $A(r)=f^{-1/2}\left(1-\frac{r_{H}}{r}\right)$, $B(r)=f^{1/2}\left(1-\frac{r_{H}}{r}\right)^{-1}$ and $C(r)=r^{2}f^{1/2}$, we can calculate $V(r)$ using (\ref{pot}) with 
$\Lambda_{\ell}$ given by:
\beq
\Lambda_{\ell}=\frac{f^{-1/2}}{r}\sqrt{1-\frac{r_{H}}{r}}\left(\ell+1\right).
\eeq
In Figure (\ref{potencialesdirac}) we show the effective potential for massless Dirac perturbations
 in four dimensional stringy black holes for various multipole numbers
$\ell$. In this Figure distances are measured in units of the black hole horizon radius $r_{H}$. We see that $V(r)$ has the form of a definite positive
potential barrier, i.e, it is a well behaved function that goes to
zero at spatial infinity and gets a maximum value at a well defined peak near the event horizon. Then we can expect that the stringy black holes are stable under massless Dirac
perturbations, a fact supported by the numerical results that we will present in the next section.

\section{Time evolution of Dirac perturbations} 
In order to integrate the equation (\ref{tevol1}) numerically we use the technique developed by Gundlach, Price and Pullin \cite{gpp}.
%%%%%%%%%%
\begin{figure}[htb!]
\begin{center}
%\resizebox{0.9\columnwidth}{!}
%{
%\includegraphics[height=16cm,width=10cm,angle=270]{grade.eps}
\includegraphics[height=8cm,width=10cm]{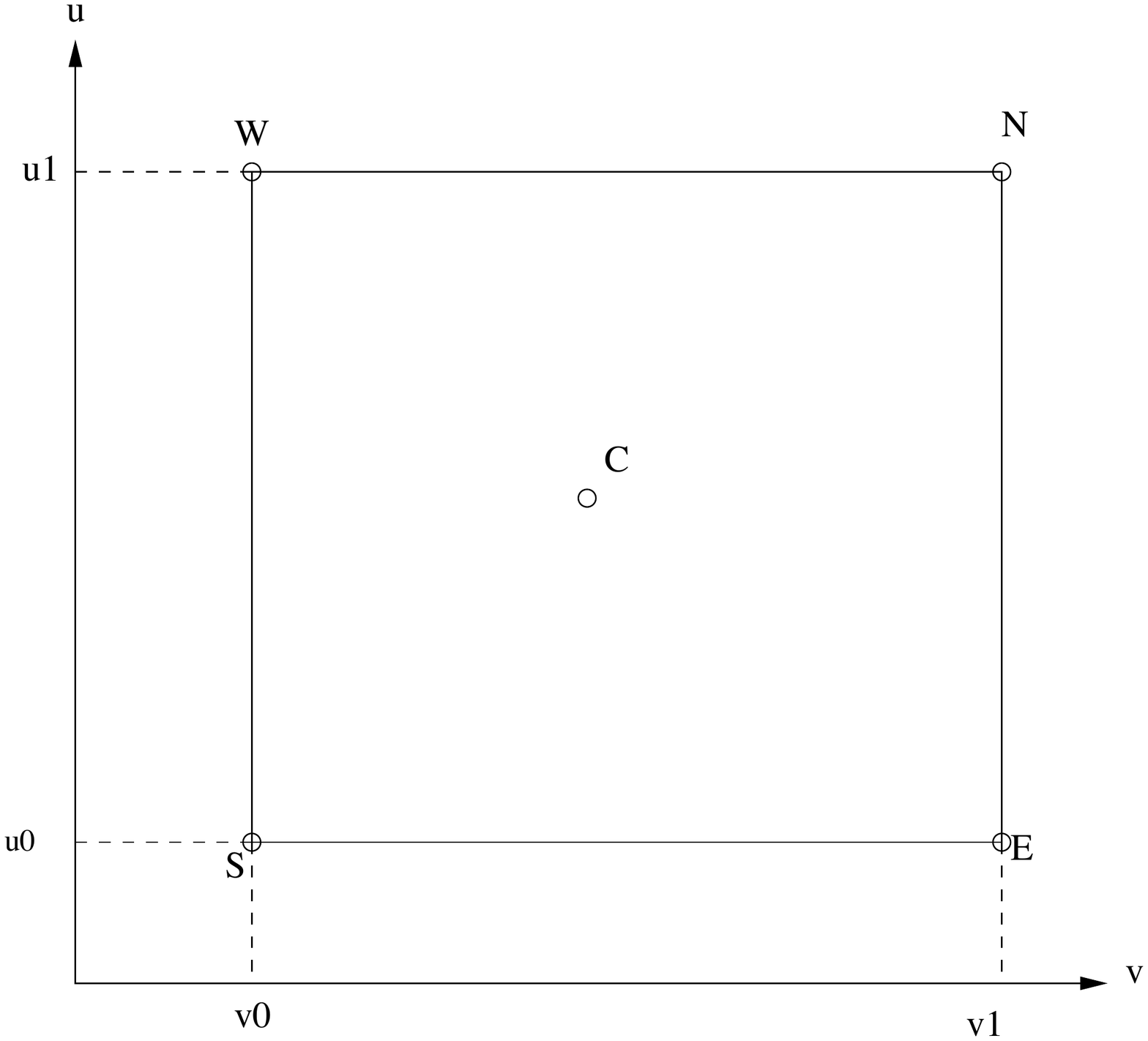}
%}
\caption{A cell of the integration grid in the plane $(u,v)$ limited by the points $N$, $S$, $E$, $W$ e $C$. This cell represents an integration step, and the initial data are specified on the left and bottom sides of the rhombus.}
\label{grade}
\end{center}
\end{figure}
%%%%%%%%%%
The first step is to rewrite the wave-like equation (\ref{tevol1}) in terms light-cone coordinates $du = dt - dr_\star$ and $dv = dt + dr_\star$
\begin{equation}\label{light-cone}
\left(4\frac{\partial^2}{\partial u\partial v}+V(u,v)\right)\zeta_{\ell}(u,v)=0.
\end{equation}
and use the following discretized version of the above equation
\begin{equation}\label{discretizacion}
\zeta_{\ell}(N)=\zeta_{\ell}(W)+\zeta_{\ell}(E)-\zeta_{\ell}(S)-\frac{\Delta u\Delta v}{8}V(S)\left(\zeta_{\ell}(W)+\zeta_{\ell}(E)\right) + \mathcal{O}(h^4),
\end{equation}
where the letters $S, W, E, N$ are used to mark the points that limits a particular integration cell of the the grid according to: $S=(u,v)$, $W=(u+\Delta u,v)$, $E=(u,v+\Delta v)$, $N=(u+\Delta u,v+\Delta v)$. In figure \ref{grade} we show the cell of a given integration step.
We see that the field value at point $N$ depends only of the field values at the points $S$, $E$ and $W$. Given a set of initial conditions 
at the two null surfaces $u=u_{0}$ and $v=v_{0}$, we can find, using (\ref{discretizacion}), the value of the field $\zeta_{\ell}$ inside the rhombus which is built on this two null surfaces. Then, by iteration of the integration cell, we find the complete data describing the evolution of the fields with time.

\begin{figure}[htb!]
\begin{center}
%\mbox{\subfigure{\includegraphics[width=3.2in]{1perfil_real_2_stringy_ro=1_a=1_l=0.eps}
\mbox{\subfigure{\includegraphics[width=3.2in]{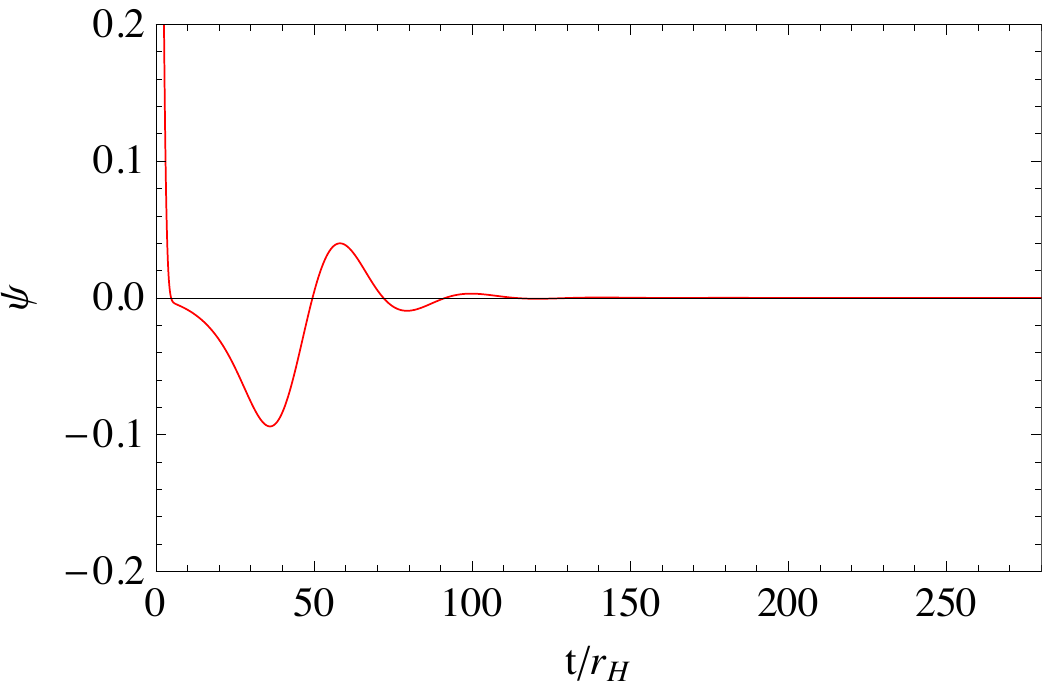}
\quad
%\subfigure{\includegraphics[width=3.2in]{perfilLog_stringy_a=1_ro=1_l=0.eps} }}}
\subfigure{\includegraphics[width=3.2in]{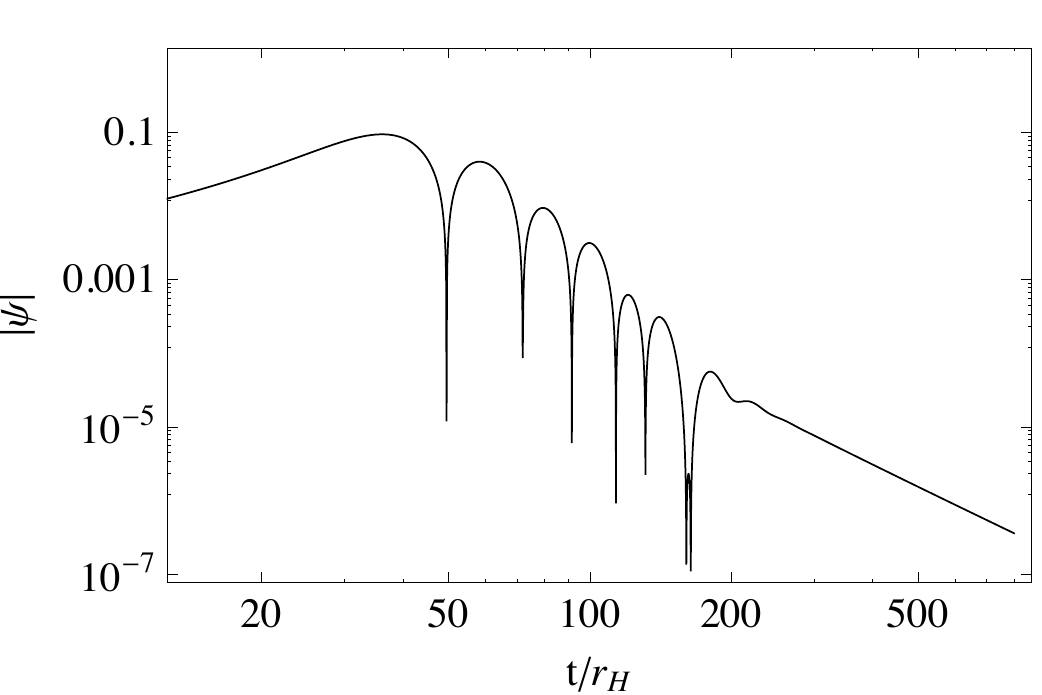} }}}
\caption{\it Normal (left) and logaritmic (right) plots of the time-domain evolution of $\ell=0$ massless Dirac perturbations.}
\label{perfil1}
\end{center}
\end{figure}

\begin{figure}[htb!]
\begin{center}
%\mbox{\subfigure{\includegraphics[width=3.2in]{1perfil_real_2_stringy_ro=1_a=1_l=1.eps}
\mbox{\subfigure{\includegraphics[width=3.2in]{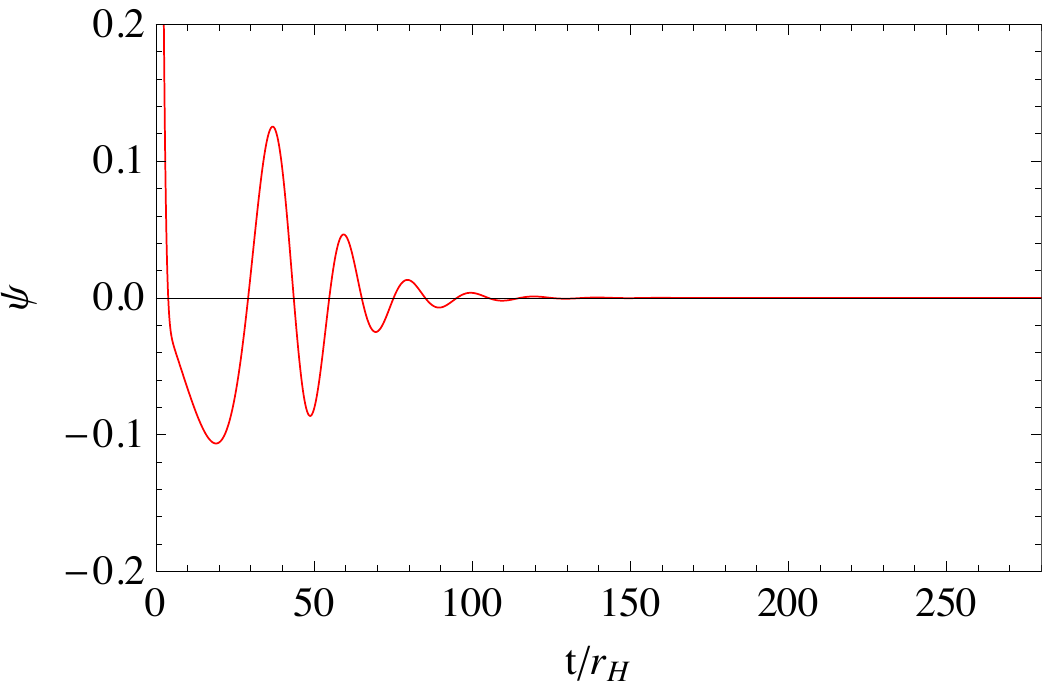}
\quad
%\subfigure{\includegraphics[width=3.2in]{perfilLog_stringy_a=1_ro=1_l=1.eps} }}}
\subfigure{\includegraphics[width=3.2in]{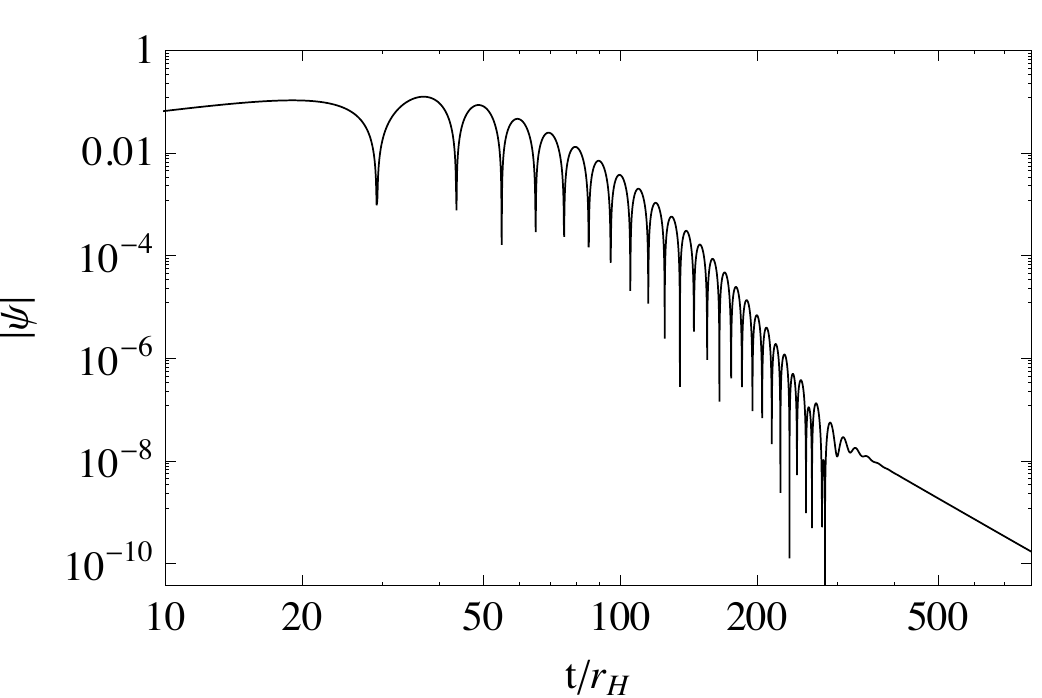} }}}
\caption{\it Normal (left) and logaritmic (right) plots of the time-domain evolution of $\ell=1$ massless Dirac perturbations.}
\label{perfil2}
\end{center}
\end{figure}

\begin{figure}[htb!]
\begin{center}
%\mbox{\subfigure{\includegraphics[width=3.2in]{perfil_real_2_stringy_ro=1_a=1_l=2.eps}
\mbox{\subfigure{\includegraphics[width=3.2in]{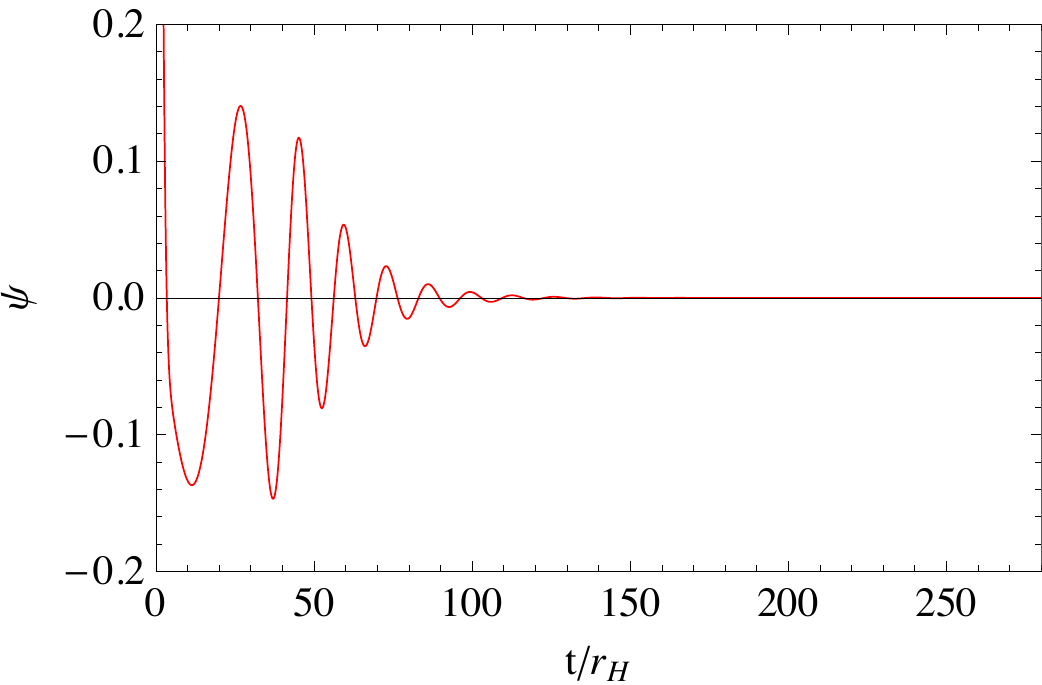}
\quad
%\subfigure{\includegraphics[width=3.2in]{perfilLog_stringy_a=1_ro=1_l=2.eps} }}}
\subfigure{\includegraphics[width=3.2in]{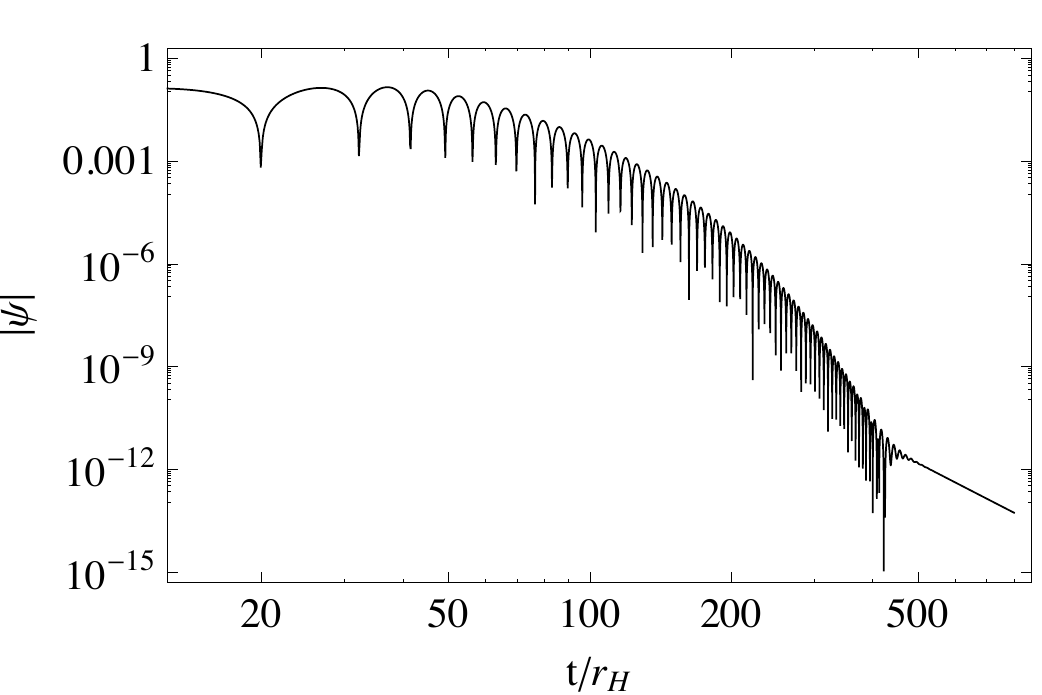} }}}
\caption{\it Normal (left) and logaritmic (right) plots of the time-domain evolution of $\ell=2$ massless Dirac perturbations.}
\label{perfil3}
\end{center}
\end{figure}

\begin{figure}[htb!]
\begin{center}
%\mbox{\subfigure{\includegraphics[width=3.2in]{perfil_real_2_stringy_ro=1_a=1_l=3.eps}
\mbox{\subfigure{\includegraphics[width=3.2in]{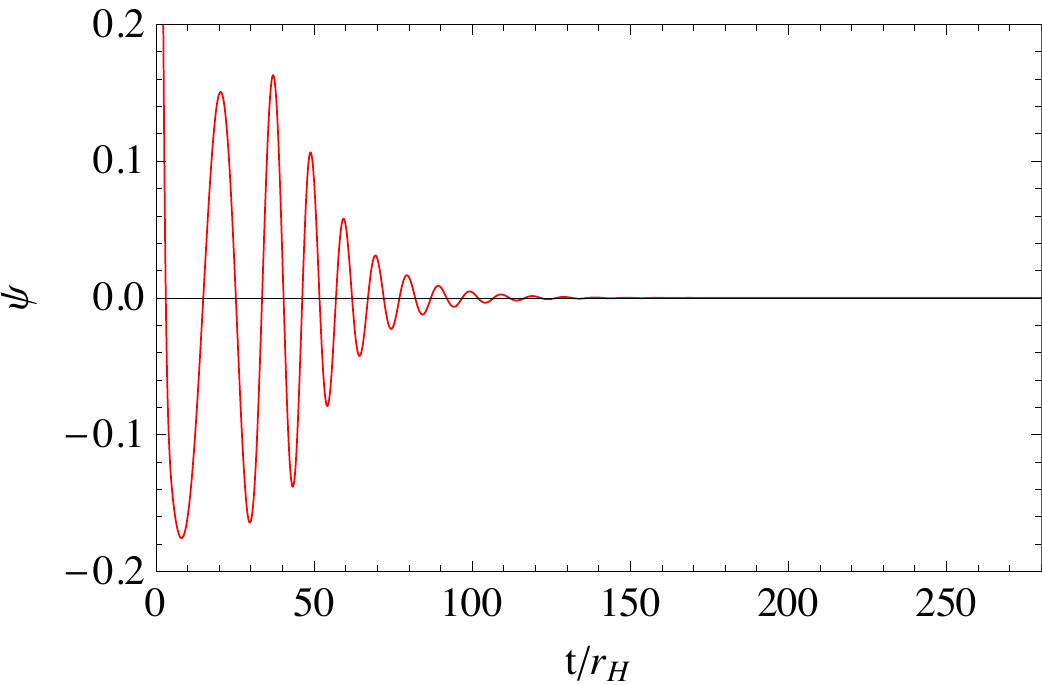}
\quad
%\subfigure{\includegraphics[width=3.2in]{perfilLog_stringy_a=1_ro=1_l=3.eps} }}}
\subfigure{\includegraphics[width=3.2in]{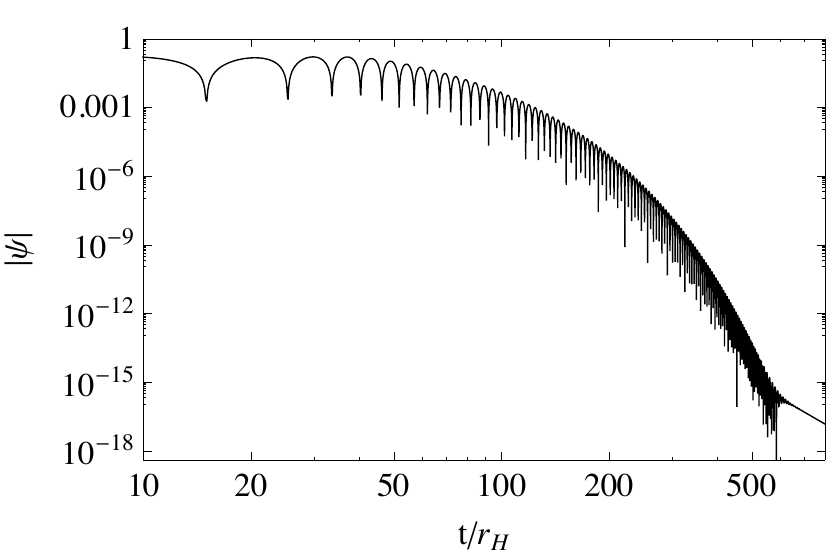} }}}
\caption{\it Normal (left) and logaritmic (right) plots of the time-domain evolution of $\ell=3$ massless Dirac perturbations.}
\label{perfil4}
\end{center}
\end{figure}

\begin{figure}[htb!]
\begin{center}
%\mbox{\subfigure{\includegraphics[width=3.2in]{1perfil_real_1_stringy_ro=1_a=1_l=6.eps}
\mbox{\subfigure{\includegraphics[width=3.2in]{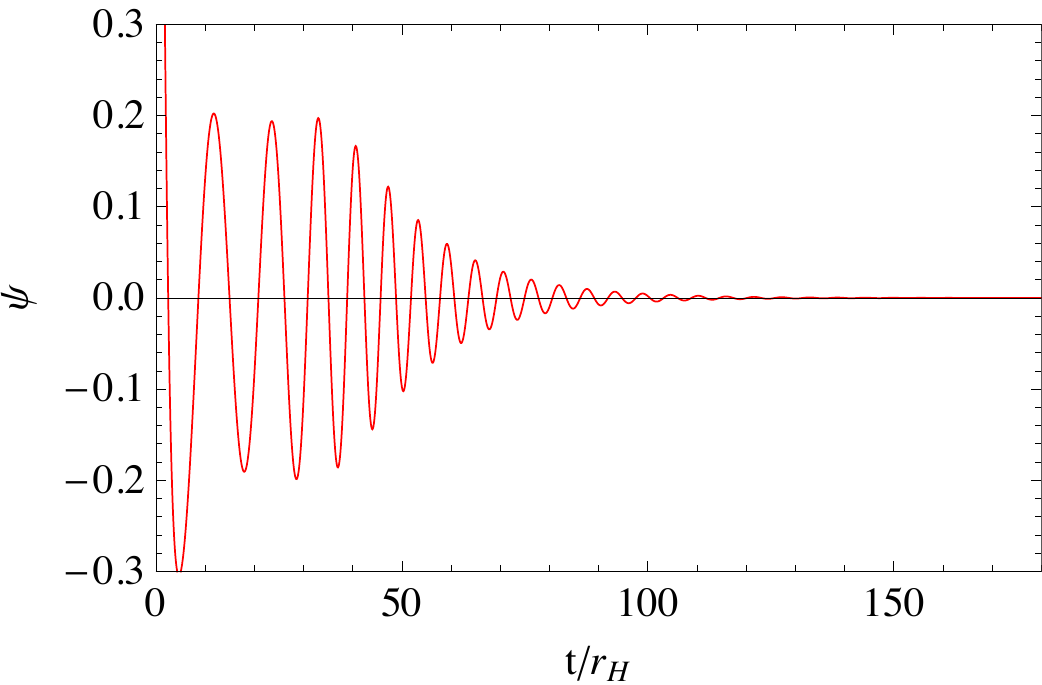}
\quad
%\subfigure{\includegraphics[width=3.2in]{2perfilLog_stringy_a=1_ro=1_l=6.eps} }}}
\subfigure{\includegraphics[width=3.2in]{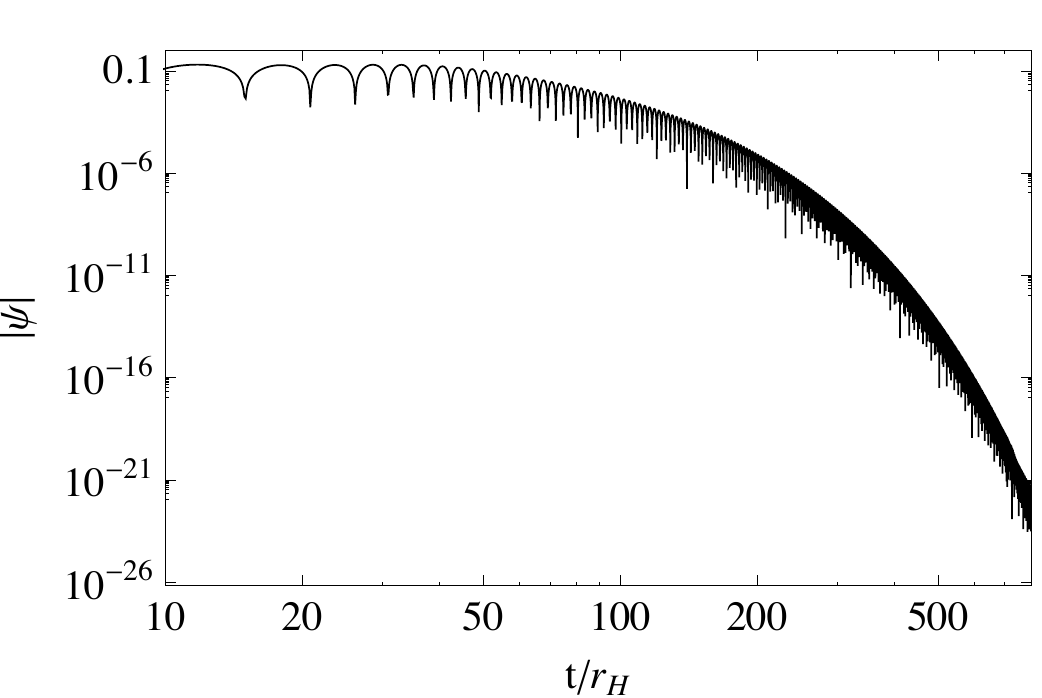} }}}
\caption{\it Normal (left) and logaritmic (right) plots of the time-domain evolution of $\ell=6$ massless Dirac perturbations.}
\label{perfil5}
\end{center}
\end{figure}

The obtained results from the integration in the case of massless Dirac fields in stringy black hole background can be observed as the time-domain profile  showed in figures (\ref{perfil1}) to (\ref{perfil5}). 
In such profiles $r=3 r_{H}$  and  the time is measured in units of black hole event horizon.
%%%%%%%%%%%%%

%%%%%%%%%%%%

%%%%%%%%%%%%

%%%%%%%%%%%%

%%%%%%%%%%%%%

As we can see, the temporal evolution of Dirac perturbations follows the usual dynamics for fields in black holes spacetimes. After a first transient stage strongly dependent on the initial conditions  and the point where the wave profile is observed, we see the characteristic exponential damping of the perturbations called quasinormal ringing, followed by a so-called power law tails at asymptotically late times. 

It is important to mention that we perform and extensive numerical exploration of the perturbative dynamics for different values
of angular number $\ell$, and we have not found any instability for massless Dirac perturbations in four dimensional stringy black holes. It is an expected result due to the positive definite character of the effective potential, as we mentioned in the previous section.

%%%%%%%%%%%%%%%%%%%%%%%%%%%%%%%%%%%%%%
\section{QNMs Using Sixth Order WKB Method and Prony fitting of characteristic data }

In the following we will assume for the function $\zeta_{\ell}(t,r)$ the time dependence:
\begin{eqnarray}
\zeta_{\ell}(t,r)=Z_{\ell}(r)\exp(-i\om_{\ell}t), 
\end{eqnarray}
Then, the function $Z_{\ell}(r)$  satisfy the Schrodinger-type equation:
\begin{equation}\label{finaleq}
\frac{d^{2} Z_{\ell}}{d r_{*}^{2}}+\left[\om^{2}- V(r)\right] Z_{\ell}(r)=0.
\end{equation}
%and 
%\begin{equation}\label{finaleq2}
%\frac{d^{2}H_{\ell}}{d r_{*}^{2}}+\left[\om^{2}- V_{-}(r)\right]H_{\ell}(r)=0
%\end{equation}
The quasinormal modes are solutions of the wave equation (\ref{tevol1})
with the specific boundary conditions requiring pure out-going waves
at spatial infinity and pure in-coming waves on the event horizon.
Thus no waves come from infinity or the event horizon. 

In order to evaluate the quasinormal modes we used two different
methods. The first is a semianalytical method to solve equation
(\ref{finaleq}) with the required boundary conditions, based in a
WKB-type approximation, that can give accurate values of the lowest
( that is longer lived ) quasinormal frequencies, and  was used in
several papers for the determination of quasinormal frequencies in a
variety of systems \cite{nuestro,splitfermion,shutz-will,iyer-will,konoplya1,konoplya2,WKB6papers}.

\begin{table}[htb!]
   % \tbl{\it Gravitino quasinormal frequencies in Schwarzschild black hole  \(\omega M\), measured in units of the black hole mass $M$ (third and sixth order WKB approximation and Prony fitting of time domain data).}
   % tamanho da fonte
   \setlength{\arrayrulewidth}{3\arrayrulewidth}  % espessura da  linha
   \setlength{\belowcaptionskip}{5pt}  % espaço entre caption e tabela
  {\begin{tabular}{|c|c|c|c|}
      \hline
      \hline
      \multicolumn{1}{|c|}{\ \ $\ell$ \ \ }& \multicolumn{1}{|c|}{\ \ $n$ \ \ }&\multicolumn{1}{|c|}{\ \ Sixth order WKB \ \ }&\multicolumn{1}{|c|}{\ \ \ \ \ \ Prony \ \ \ \ \ \ } \\
      \hline
      $0$ & $0$  & $0.1527-0.0623i$&$ \ \ 0.1525-0.0620i \ \ $ \\
      \hline
  $1$ & $0$ &  $0,3147-0.0628i$&$ 0,3147-0.0626i$ \\
      \hline
  $2$ & $0$ &  $0.4744-0.0629i$&$ 0.4744-0.0629i$ \\
      \hline
 $2$ & $1$ & $0.4663-0.1899i$&$ 0.4661-0.1896i$ \\
     \hline
     $3$ & $0$& $0.6337-0.0629i$&$ 0.6337-0.0629i$ \\
      \hline
 $3$ & $1$ &  $0.6275-0.1894i$&$0.6271-0.1880 i$ \\
      \hline
$3$ & $2$ &  $0.6156-0.3180i$&$ -$ \\
     \hline
       $4$ &  $0$ & $0.7928-0.0629i$&$ 0.7928-0.0629i $ \\
      \hline
 $4$ & $1$ &  $0.7878-0.1892i$&$0.7878-0.1892i$ \\
      \hline
$4$ & $2$ &  $0.7781-0.3169i$&$ -$ \\
     \hline
      $4$ & $3$ &$ 0.7641-0.4466i$& $-$\\
      \hline
       $5$ & $0$  & $0.9518-0.0629i$&$ 0.9518-0.0629i $ \\
      \hline
 $5$ & $1$ &  $0.9476-0.1891i$&$ 0.9476-0.1891i $ \\
     \hline
     $5$ & $2$ &$0.9394-0.3162i$&$-$ \\
\hline
$5$ & $3$ &  $0.9275-0.4448i$&$ - $ \\
\hline
$5$ & $4$ &  $0.9121-0.5755i$&$ - $ \\
     \hline
       \hline
   \end{tabular}\label{frecuenciasdirac1}}
\caption{\it Dirac quasinormal frequencies \(\omega r_{H}\) for $\ell=0$ to $\ell=5$.}
      \end{table}
\begin{table}[htb!]
   % \tbl{\it Gravitino quasinormal frequencies in Schwarzschild black hole  \(\omega M\), measured in units of the black hole mass $M$ (third and sixth order WKB approximation and Prony fitting of time domain data).}
   % tamanho da fonte
   \setlength{\arrayrulewidth}{3\arrayrulewidth}  % espessura da  linha
   \setlength{\belowcaptionskip}{5pt}  % espaço entre caption e tabela
  {\begin{tabular}{|c|c|c|c|}
      \hline
      \hline
      \multicolumn{1}{|c|}{\ \ $\ell$ \ \ }& \multicolumn{1}{|c|}{\ \ $n$ \ \ }&\multicolumn{1}{|c|}{\ \ Sixth order WKB \ \ }&\multicolumn{1}{|c|}{\ \ \ \ \ \ Prony \ \ \ \ \ \ } \\
      \hline
      $6$ & $0$  & $1.1107-0.0629i$ &$ 1.1108-0.0629i  $ \\
      \hline
  $6$ & $1$ &  $1.1071-0.1890i$&$ 1.1071-0.1890i$ \\
      \hline
  $6$ & $2$ &  $1.100-0.3158i$&$ -$ \\
      \hline
 $6$ & $3$ & $1.0897-0.4437i$&$ -$ \\
     \hline
 $6$ & $4$ &  $1.0762-0.5732i$&$ -$ \\
      \hline
$6$ & $5$ &  $1.0600-0.7045i$&$ -$ \\
     \hline
       $7$ &  $0$ & $1.2696-0.0629i$&$ 1.2697-0.0629i $ \\
      \hline
 $7$ & $1$ &  $1.2665-0.1890$&$1.2665-0.1889i$ \\
      \hline
$7$ & $2$ &  $1.2603-0.3155i$&$ -$ \\
     \hline
      $7$ & $3$ &$ 1.2511-0.4430i$& $-$\\
      \hline
       $7$ & $4$  & $1.2392-0.5716i$&$ - $ \\
      \hline
 $7$ & $5$ &  $1.2246-0.7018i$&$ - $ \\
     \hline
     $7$ & $6$ &$1.2077-0.8336i$&$-$ \\
     \hline
       \hline
   \end{tabular}\label{frecuenciasdirac2}}
\caption{\it Dirac quasinormal frequencies   \(\omega r_{H}\) for $\ell=6$ and $\ell=7$ .}
      \end{table}
\begin{table}[htb!]
   % \tbl{\it Gravitino quasinormal frequencies in Schwarzschild black hole  \(\omega M\), measured in units of the black hole mass $M$ (third and sixth order WKB approximation and Prony fitting of time domain data).}
   % tamanho da fonte
   \setlength{\arrayrulewidth}{3\arrayrulewidth}  % espessura da  linha
   \setlength{\belowcaptionskip}{5pt}  % espaço entre caption e tabela
   {\begin{tabular}{|c|c|c|c|}
      \hline
      \hline
      \multicolumn{1}{|c|}{\ \ $\ell$ \ \ }& \multicolumn{1}{|c|}{\ \ $n$ \ \ }&\multicolumn{1}{|c|}{\ \ Sixth order WKB \ \ }&\multicolumn{1}{|c|}{\ \ \ \ \ \ Prony \ \ \ \ \ \ } \\
      \hline
      $8$ & $0$  & $1.4285-0.0629i$ &$ 1.4287-0.0629i $ \\
      \hline
  $8$ & $1$ &  $1.4257-0.1889i$&$ 1.4258-0.1888i$ \\
      \hline
  $8$ & $2$ &  $1.4202-0.3154i$&$ -$ \\
      \hline
 $8$ & $3$ & $1.4120-0.4425i$&$ -$ \\
     \hline
     $8$ & $4$& $1.4012-0.5706i$&$ - $ \\
      \hline
 $8$ & $5$ &  $1.3881-0.6999i$&$ -$ \\
      \hline
$8$ & $6$ &  $1.3727-0.8305i$&$ -$ \\
     \hline
       $8$ &  $7$ & $1.3554-0.9628i$&$ - $ \\
      \hline
        \hline
   \end{tabular}\label{frecuenciasdirac3}}
\caption{\it Dirac quasinormal frequencies \(\omega r_{H}\) for $\ell=8$.}
      \end{table}
\begin{table}[htb!]
   % \tbl{\it Gravitino quasinormal frequencies in Schwarzschild black hole  \(\omega M\), measured in units of the black hole mass $M$ (third and sixth order WKB approximation and Prony fitting of time domain data).}
   % tamanho da fonte
   \setlength{\arrayrulewidth}{3\arrayrulewidth}  % espessura da  linha
   \setlength{\belowcaptionskip}{5pt}  % espaço entre caption e tabela
   {\begin{tabular}{|c|c|c|c|}
      \hline
      \hline
      \multicolumn{1}{|c|}{\ \ $\ell$ \ \ }& \multicolumn{1}{|c|}{\ \ $n$ \ \ }&\multicolumn{1}{|c|}{\ \ Sixth order WKB \ \ }&\multicolumn{1}{|c|}{\ \ \ \ \ \ Prony \ \ \ \ \ \ } \\
      \hline
      $9$ & $0$  & $1.5873-0.0629i$ &$ 1.5876-0.0629i $ \\
      \hline
  $9$ & $1$ &  $1.5848-0.1889i$&$ 1.5850-0.1888i$ \\
      \hline
  $9$ & $2$ &  $1.5798-0.3152i$&$ -$ \\
      \hline
 $9$ & $3$ & $1.5724-0.4421i$&$ -$ \\
     \hline
     $9$ & $4$& $1.5627-0.5698i$&$ - $ \\
      \hline
 $9$ & $5$ &  $1.5507-0.6985i$&$ -$ \\
      \hline
$9$ & $6$ &  $1.5367-0.8283i$&$ -$ \\
     \hline
       $9$ &  $7$ & $1.5207-0.9594i$&$ - $ \\
      \hline
$9$ &  $8$ & $1.5030-1.0920i$&$ - $ \\
      \hline
        \hline
   \end{tabular}\label{frecuenciasdirac3}}
\caption{\it Dirac quasinormal frequencies  \(\omega r_{H}\) for $\ell=9$.}
      \end{table}
\begin{table}[htb!]
   % \tbl{\it Gravitino quasinormal frequencies in Schwarzschild black hole  \(\omega M\), measured in units of the black hole mass $M$ (third and sixth order WKB approximation and Prony fitting of time domain data).}
   % tamanho da fonte
   \setlength{\arrayrulewidth}{3\arrayrulewidth}  % espessura da  linha
   \setlength{\belowcaptionskip}{5pt}  % espaço entre caption e tabela
   {\begin{tabular}{|c|c|c|c|}
      \hline
      \hline
      \multicolumn{1}{|c|}{\ \ $\ell$ \ \ }& \multicolumn{1}{|c|}{\ \ $n$ \ \ }&\multicolumn{1}{|c|}{\ \ Sixth order WKB \ \ }&\multicolumn{1}{|c|}{\ \ \ \ \ \ Prony \ \ \ \ \ \ } \\
      \hline
      $10$ & $0$  & $1.7462-0.0629i$ &$ 1.7466-0.0629i $ \\
      \hline
  $10$ & $1$ &  $1.7439-0.1889i$&$1.7442-0.1887i $ \\
      \hline
  $10$ & $2$ &  $1.7393-0.3151i$&$ -$ \\
      \hline
 $10$ & $3$ & $1.7326-0.4419i$&$ -$ \\
     \hline
     $10$ & $4$& $1.7237-0.5692i$&$ - $ \\
      \hline
 $10$ & $5$ &  $1.7127-0.6974i$&$ -$ \\
      \hline
$10$ & $6$ &  $1.6998-0.8266i$&$ -$ \\
     \hline
       $10$ &  $7$ & $1.6850-0.9569i$&$ - $ \\
      \hline
 $10$ &  $8$ & $1.6686-1.0884i$&$ - $ \\
      \hline
$10$ &  $9$ & $1.6505-1.2212i$&$ - $ \\
      \hline
        \hline
   \end{tabular}\label{frecuenciasdirac3}}
\caption{\it Dirac quasinormal frequencies \(\omega r_{H}\) for $\ell=10$ .}
      \end{table}
\begin{figure}[!]
\begin{center}
%\scalebox{0.55}{\includegraphics{qnm_re_vs_im_stringy_dirac.eps}}
\scalebox{0.55}{\includegraphics{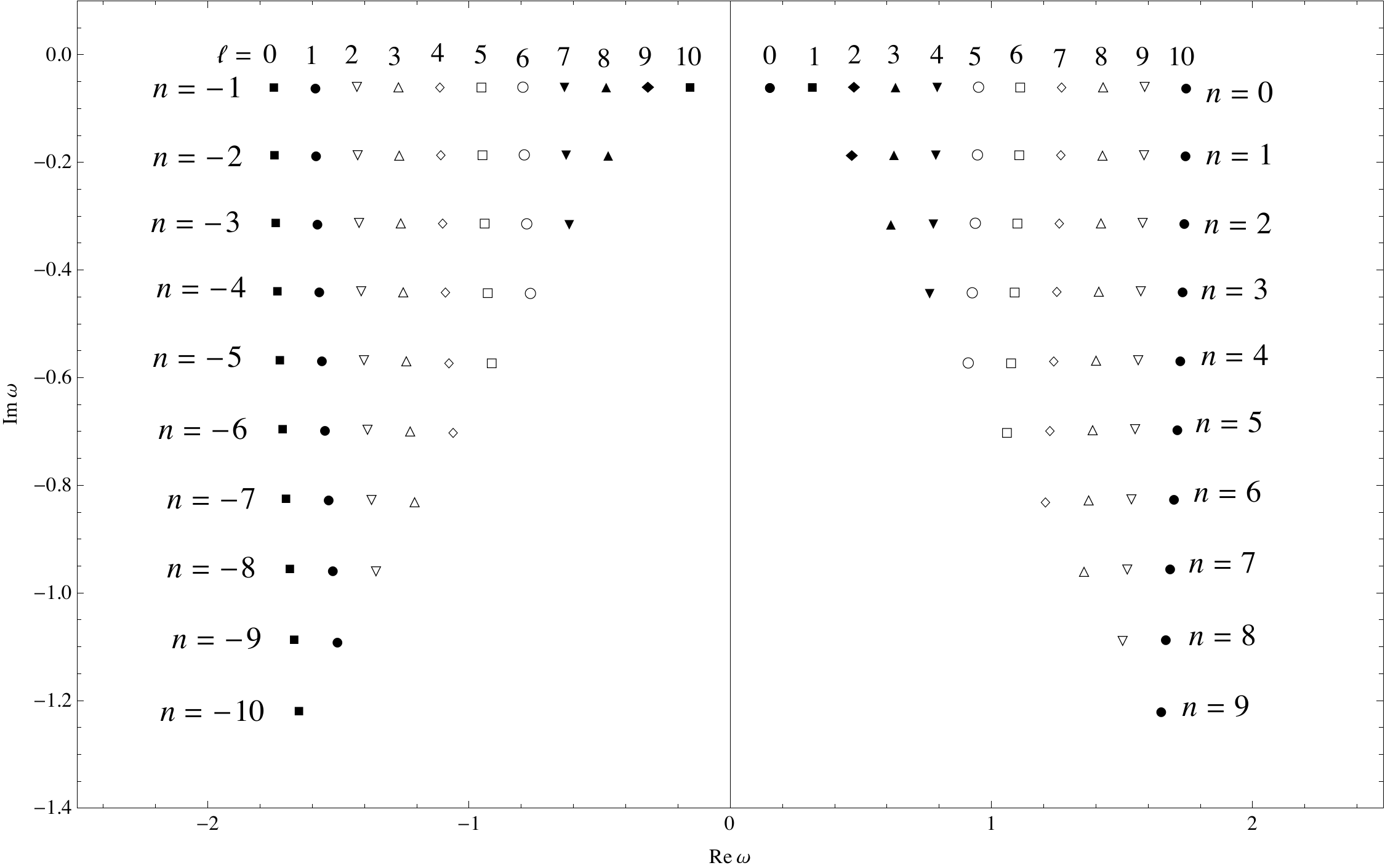}}
\caption{Massles Dirac quasinormal frequencies for various values of $l$ e $n$.}
\end{center}
\label{QNM}
\end{figure}
The WKB technique was applied
up to first order to finding quasinormal modes for the first time by
Shutz and Will \cite{shutz-will}. Latter this approach was extended
to the third order beyond the eikonal approximation by Iyer and Will
\cite{iyer-will} and to the sixth order by Konoplya
\cite{konoplya1,konoplya2}. We use in our numerical calculation of quasinormal
modes this sixth order WKB expansion. The sixth order WKB expansion
gives a relative error which is about two orders less than the third
WKB order, and allows us to determine the quasinormal frequencies
through the formula
\begin{equation}\label{WKB6}
    i\frac{\left(\omega^{2}-V_{0}\right)}{\sqrt{-2V_{0}^{''}}}-\sum_{j=2}^{6}\Pi_{j}=n+\frac{1}{2},
\end{equation}
where $n= 0,1,2,...$ if $Re(\om)>0$ or $n= -1,-2,-3,..$ if $ Re(\om)<0$ is the overtone number. In (\ref{WKB6}) \(V_{0}\) is the value of the potential at its maximum as a
function of the tortoise coordinate, and \(V_{0}^{''}\) represents
the second derivative of the potential with respect to the tortoise
coordinate at its peak. The correction terms \(\Pi_{j}\) depend on
the value of the effective potential and its derivatives ( up to the
2i-th order) in the maximum, see \cite{zhidenkothesis} and
references therein.

The second method that we used to find the quasinormal frequencies
was the Prony method \cite{berti3,zhidenkothesis} for fitting the time domain profile data by
superposition of damping exponents in the form
\begin{equation}\label{Prony}
    \psi\left(t\right)=\sum_{k=1}^{p}C_{k}e^{-iw_{k}t}.
\end{equation}
Assuming that the quasinormal ringing stage begins at $t=0$ and ends at
$t=Nh$, where $N\geq2p-1$, then the expession (\ref{Prony}) is
satisfied for each value in the time profile data
\begin{equation}\label{Prony2}
    x_{n}\equiv\psi\left(nh\right)=\sum_{k=1}^{p}C_{k}e^{-iw_{k}nh}=\sum_{k=1}^{p}C_{k}z_{k}^{n}.
\end{equation}
From the above expression, we can determine, as we know $h$, the
quasinormal frequencies $\omega_{i}$ once we have determined $z_{i}$
as functions of $x_{n}$. The Prony method allows to find the $z_{i}$
as roots of the polinomial function $A(z)$ defined as
\begin{equation}\label{Prony3}
    A\left(z\right)=\prod_{k=1}^{p}\left(z-z_{k}\right)=\sum_{m=0}^{p}\alpha_{m}z^{p-m} \
    \ , \ \ \ \ \ \ \ \ \ \ \ \alpha_{0}=1.
\end{equation}
It is possible to show that the unknown coefficients $\alpha_{m}$ of
the polinomial function $A(z)$ satisfy
\begin{equation}\label{Prony5}
    \sum_{m=1}^{p}\alpha_{m}x_{n-m}=-x_{n}.
\end{equation}
Solving the $N-p+1\geq p$ linear equations (\ref{Prony5}) for
$\alpha_{m}$ we can determine numerically the roots $z_{a}$ and then
the quasinormal frequencies.

It is important to mention the fact that with the Prony method we
can obtain very accurate results for the quasinormal frequencies,
but the practical application of the method is limited because we
need to know with precision the duration of the quasinormal ringing
epoch. As this stage is not a precisely defined time interval, in
practice, it is difficult to determine when the quasinormal ringing
begins. Therefore, we are able to calculate with high accuracy only
two or, sometimes three dominant frequencies.

Tables (I) to (V) shows the results for the quasinormal frequencies measured in units of black hole horizon radius $r_{H}$ for some values of the multipole moment $\ell$. As it is observed, the sixth order WKB
approach gives results in good correspondence with that obtained by fitting the numerical integration data using Prony technique. 

As usual, the oscillation frequency increase for higher multipole and fixed overtone numbers. The fundamental mode, i.e, that with $\ell=0$ and $n=0$ is more long-lived with respect to the other modes, but it is interesting that 
imaginary part of the $n=0$ quasinormal frequencies reach very quickly a fixed value for higher multipole numbers. 

This situation is different for higher overtone numbers, and the imaginary part for a fixed $n$ increases
with $\ell$. For $n=1$ we see that its reach a constant value from the $\ell=8$, within numerical accuracy. As we expected for stability, all quasinormal frequencies calculated in this work have a well defined negative imaginary
part.

%%%%%%%%%%%%%%%%%%%%%%%%%%%
\section{Long Time Tails}
Another important point to study is the relaxation of the perturbing fermion field outside the black hole. It is a known result that in Schwarzschild black hole neutral massless fields had a late-time behavior for a fixed $r$ dominated by a factor
$t^{-(2\ell+3)}$ for each multipole moment $\ell$ \cite{Price,burko}. 

To study the late-time behavior, we numerically fit the profile data
obtained in the appropriate region of the time domain, to extract the power law exponents that describe the relaxation. As a test of our numerical fitting scheme, we obtained the power law exponents for the massless Dirac field considered in this paper
in the space-time corresponding to four dimensional Schwarzschild black hole with unit event horizon. As we expected, the results obtained are consistent with the power law falloff mentioned in the previous paragraph. 
%%%%%%%%%%%%
%\begin{figure}[htb!]
%\begin{center}
%\mbox{\subfigure{\includegraphics[width=3.2in]{stringy-dirac-tailfitL=0}
%\quad
%\subfigure{\includegraphics[width=3.2in]{stringy-dirac-tailfitL=1} }}}
%\caption{\it Normal (left) and logaritmic (right) plots of the time-domain evolution of $\ell=6$ massless Dirac perturbations.}
%\label{perfil5}
%\end{center}
%\end{figure}
%%%%%%%%%%%%%%%%%%%%%%%

\begin{figure}[!]
%\scalebox{0.55}{\includegraphics{stringy-dirac-tailfitL=0.eps}}
\scalebox{0.70}{\includegraphics{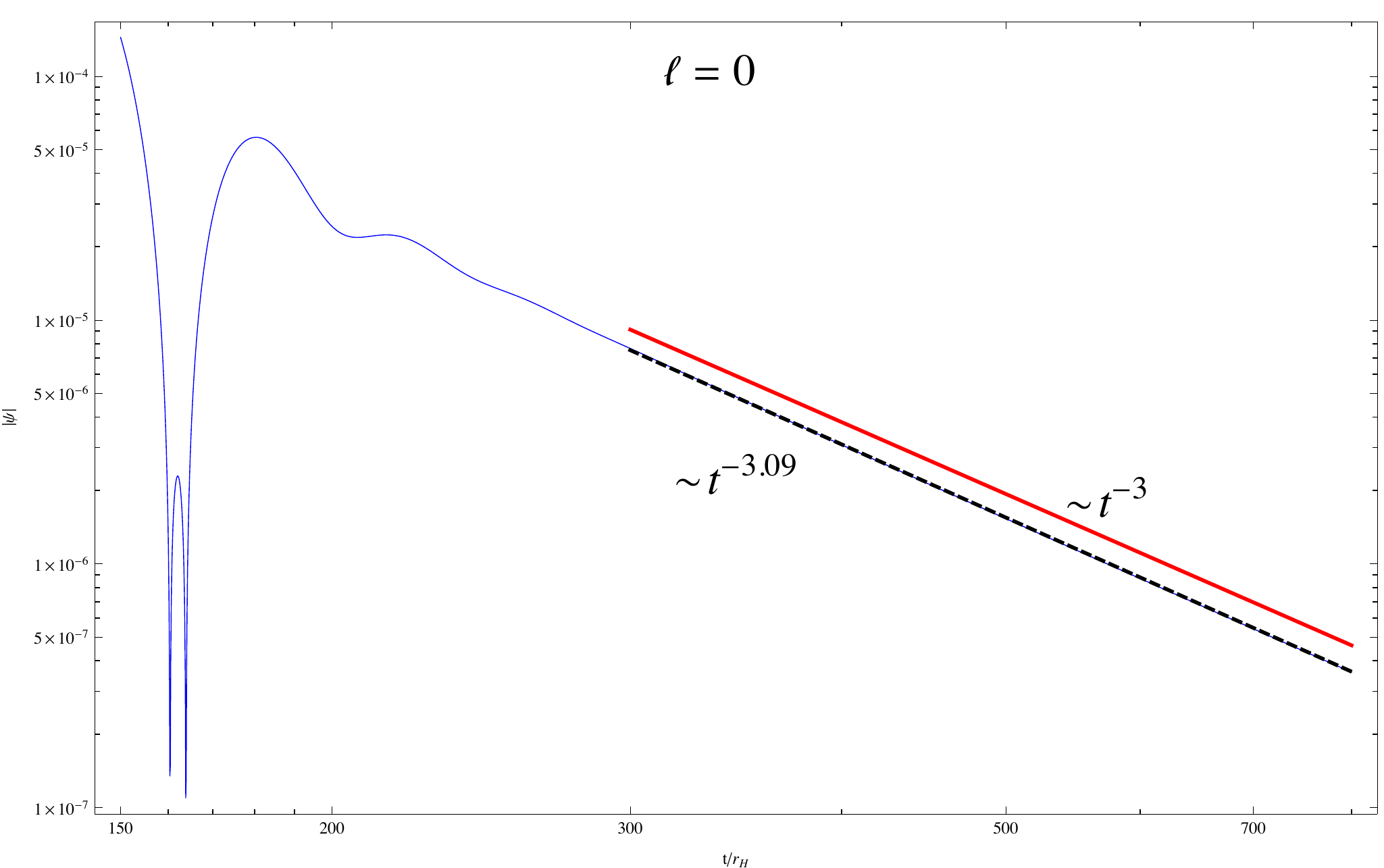}}
\caption{{\it{Tail for $\ell=0$. The power-law coefficients were estimated from numerical data represented in the dotted line. The full red line is the possible analytical result. }}}
\label{tail1}
\end{figure}
\begin{figure}[!]
%\scalebox{0.55}{\includegraphics{stringy-dirac-tailfitL=1.eps}}
\scalebox{0.55}{\includegraphics{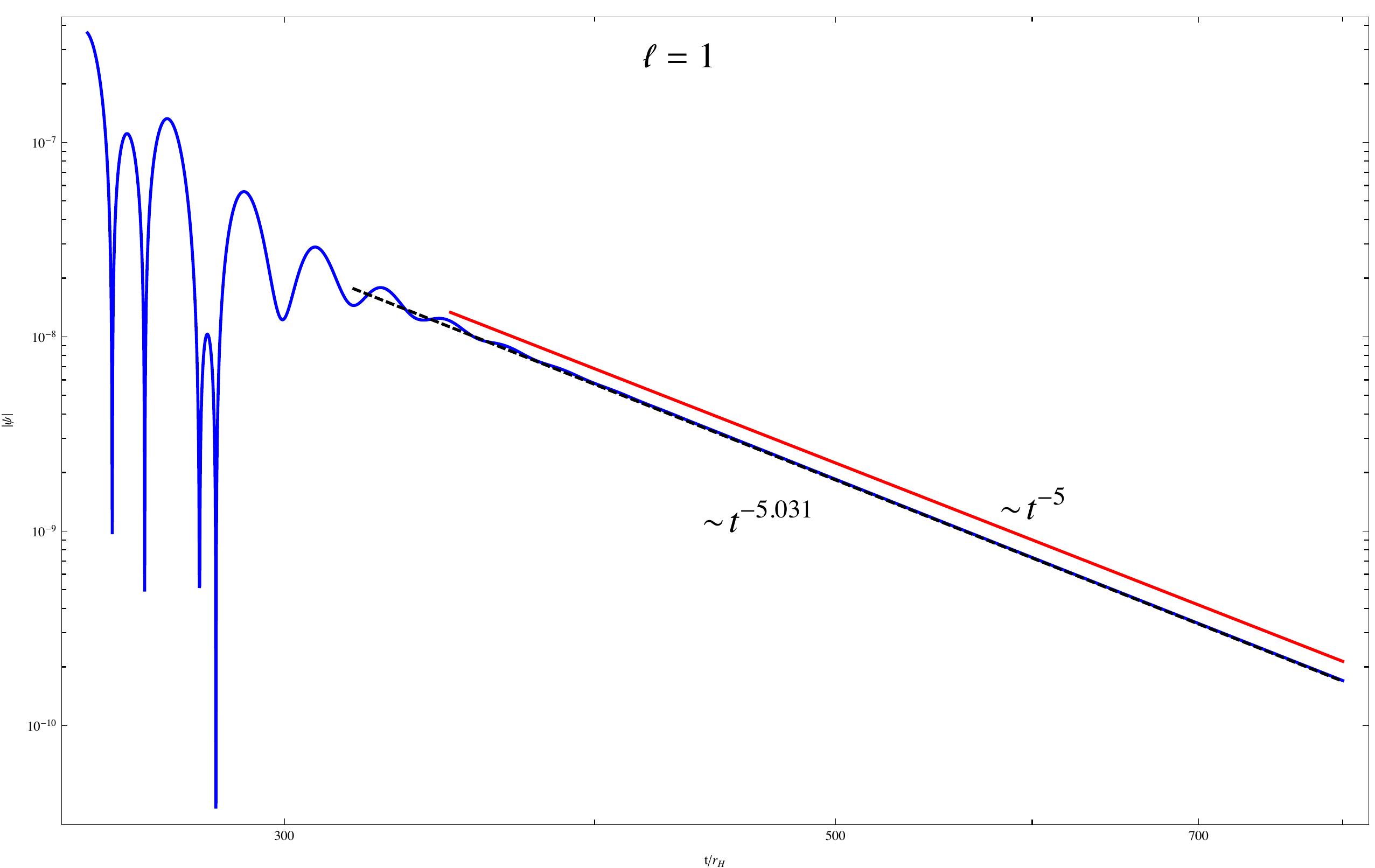}}
\caption{{\it{Tail for $\ell=1$. The power-law coefficients were estimated from numerical data represented in the dotted line. The full red line is the possible analytical result . }}}
\label{tail2}
\end{figure}
\begin{figure}[!]
%\scalebox{0.75}{\includegraphics{stringy-dirac-tailfitL=2.eps}}
\scalebox{0.75}{\includegraphics{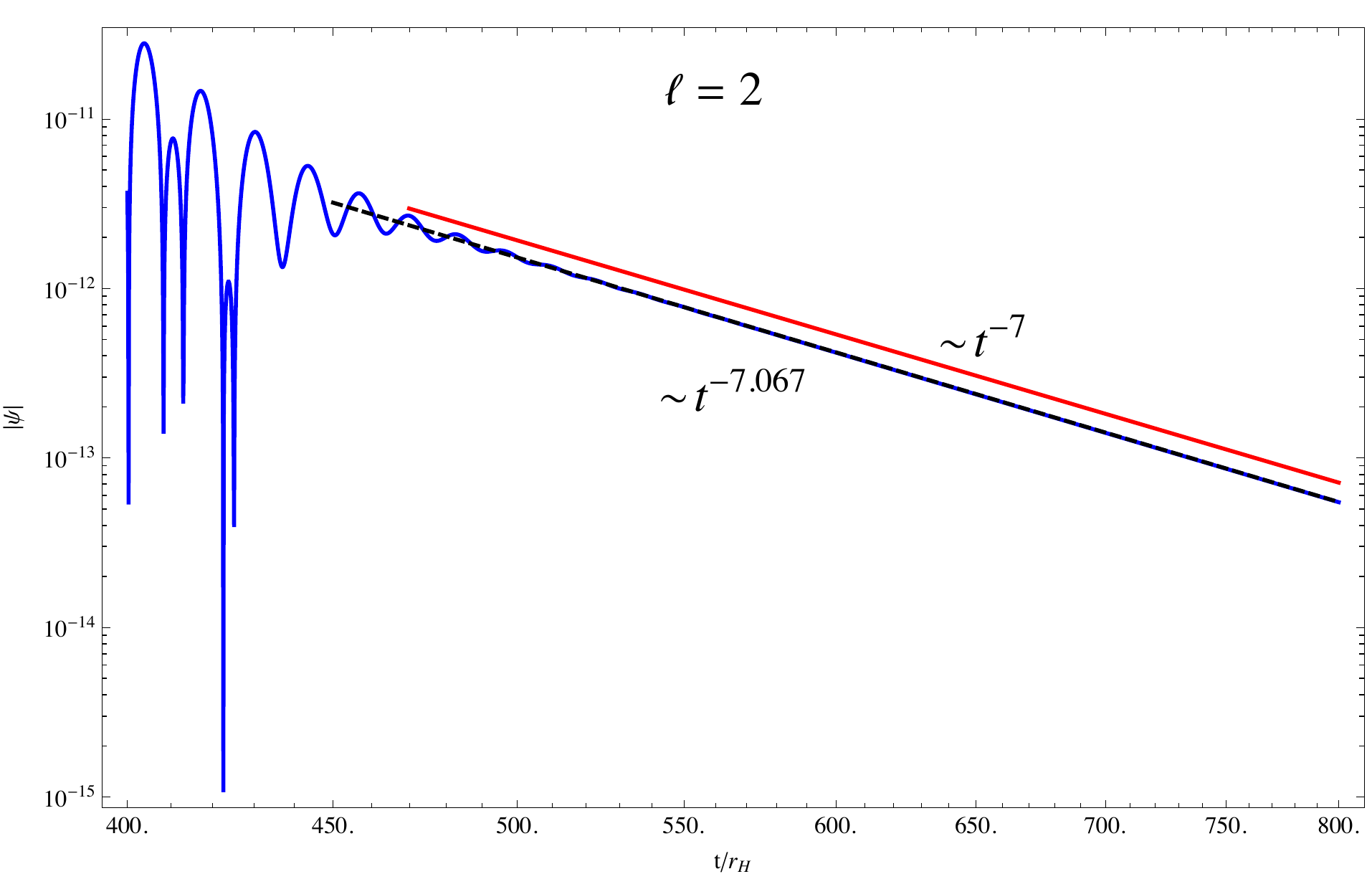}}
\caption{{\it{Tail for $\ell=2$. The power-law coefficients were estimated from numerical data represented in the dotted line. The full red line is the possible analytical result . }}}
\label{tail3}
\end{figure}
\begin{figure}[!]
%\scalebox{0.75}{\includegraphics{stringy-dirac-tailfitL=3.eps}}
\scalebox{0.80}{\includegraphics{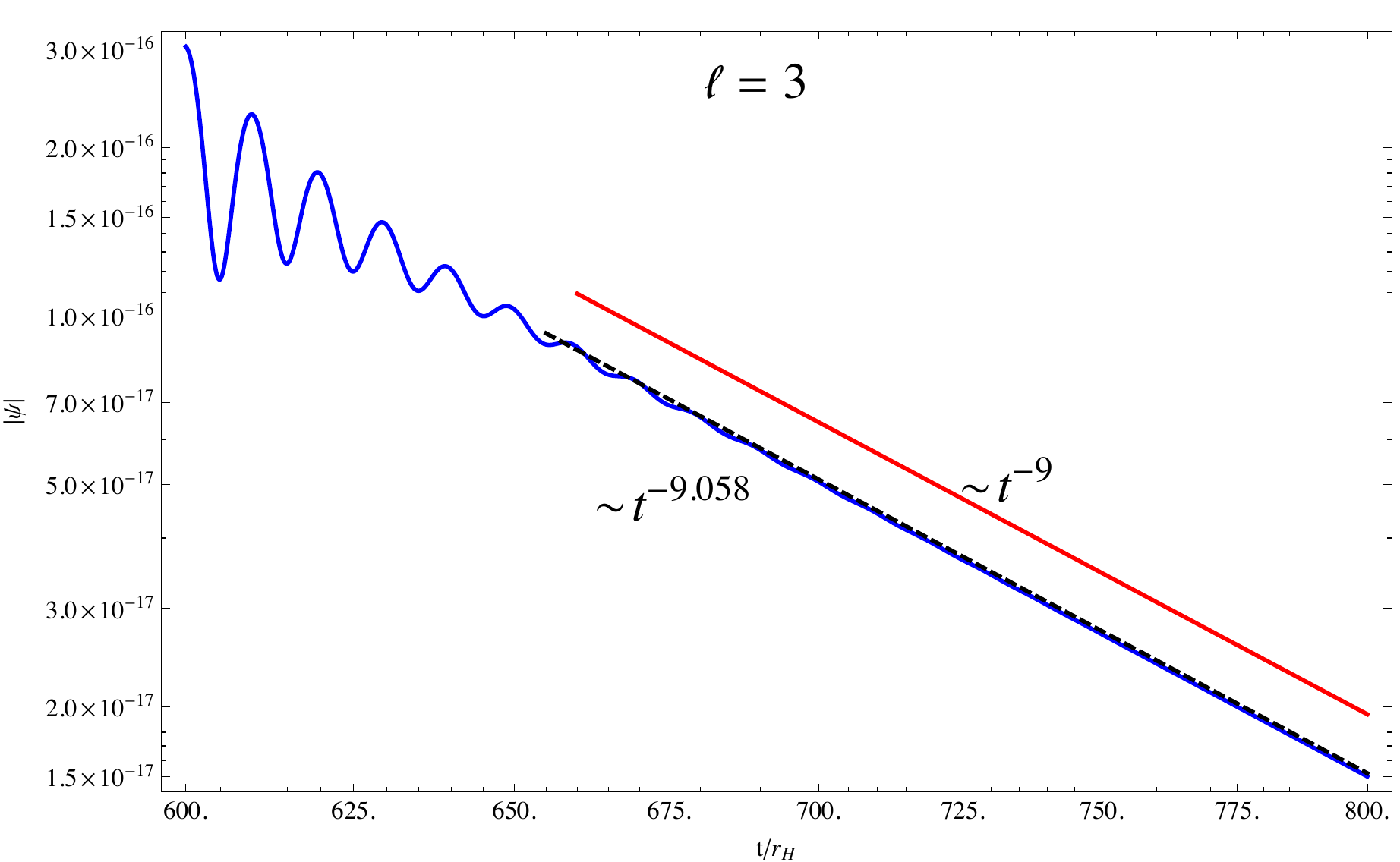}}
\caption{{\it{Tail for $\ell=3$. The power-law coefficients were estimated from numerical data represented in the dotted line. The full red line is the possible analytical result . }}}
\label{tail4}
\end{figure}
The case for the late-time fermion decay in (3+1)-stringy black hole is similar, as shows the results of our numerical fitting presented in Figures  (\ref{tail1}) to (\ref{tail4}).

For $\ell=0$ modes, the numerical decay factor is proportional to $t^{-3.09}$, for $\ell=1$ is proportional to $t^{-5.03}$. From the results displayed in Figures (\ref{tail1}) to (\ref{tail4}), and the other obtained fitting the time domain data
for all the values of angular numbers considered in our numerical work, we can determine the power law factor that dominates the falloff of the Dirac field at very late times, in the space-time of the stringy black hole. If we suppose that the decay in the Dirac case is governed by factors of the form $t^{-(\alpha\ell+\beta)}$ for each multipole moment $\ell$, then using our numerical results we obtain a decay factor for the Dirac perturbations 
at late times, that is of the form
$t^{-(2\ell+3)}$. We remark at this point that this dependance is only a result consistent with our numerical data for all values of the multipole moments studied, and remains an open problem to obtain this directly from analytical calculations. Then, we can conclude that,
outside four dimensional stringy black holes, as well as Schwarzschild black hole, the massless Dirac field shows identical decay at late times.

%%%%%%%%%%%%%%%%%%%%%%%%%%%%%%
\section{Large Angular Momentum}

Although in general the calculation of quasinormal frequencies is performed though numerical tools, we can use the WKB formula to get an analytic expression for the frequencies in the limit of large angular momentum $\ell$ .
 It can be obtained  expanding the effective potential in terms of small values of $1/\ell$ and using the lowest order of WKB approximation. In the following we suppose for simplicity $Q_{i}=Q$. So, we get

\begin{equation}\label{wkb_large_l}
\omega^{2}=\ell^{2}\sigma(r_{m})-i(n+\frac{1}{2})\sqrt{-2\frac{d^{2}\sigma(r_{m})}{dr^{2}}},
\end{equation}
where
\[
\sigma(r)=\left(1-\frac{r_{H}}{r}\right)\left(1+\frac{r_{H}Q}{r}\right)^{-4},
\]
and $r_{m}$ is where the peak of $V(r)$ occurs, and its given by
\[
r_{m}=\frac{r_{H}}{2}\left\{\left(Q+\frac{3}{2}\right)+\left[\left(Q+\frac{3}{2}\right)^{2}-2Q\right]^{\frac{1}{2}}\right\}.
\]

The quasinormal frequencies calculated using the above expression completely agree with those ones obtained through the six-order WKB approach in the large angular momentum regime. In the table (VI), 
we can see that the imaginary part of the frequencies for fixed overtone number $n=0$ goes to a constant value in the limit of large values of $\ell$. On the other hand the real part increases as $\ell$ becomes larger.

\begin{table}[htb!]
      \setlength{\arrayrulewidth}{3\arrayrulewidth}  % espessura da  linha
   \setlength{\belowcaptionskip}{5pt}  % espaço entre caption e tabela
   {\begin{tabular}{|c|c|c|}
      \hline
      \hline
      \multicolumn{1}{|c|}{\ \ $\ell$ \ \ }&\multicolumn{1}{|c|}{\ \  Dirac quasinormal frequencies\ \ } \\
      \hline
      $50$ & $3.96989-0.0314673 i$  \\
      \hline
  $100$ & $7.93959 - 0.0314681 i$ \\
      \hline
  $150$ &   $11.9093-0.0314682 i$ \\
      \hline
 $200$ & $15.8791-0.0314683 i$\\
     \hline
         \hline
   \end{tabular}\label{largel}}
\caption{\it Dirac quasinormal frequencies in the large angular momentum limit with $n=0$, $r_{H}=2$ and $Q=1$.}
      \end{table}

%%%%%%%%%%%%%%%%%%%%%%%%%%%%%%

\section{Concluding remarks}
We have studied the evolution of massless Dirac perturbations in the space-time of a (3+1)-dimensional
black hole solution from string theory. Solving numerically the
time evolution equation for this perturbations, we find similar time
domain profiles as in the case of Dirac fields in other four dimensional black hole backgrounds. At intermediary times
the evolution of fermion perturbations is dominated by quasinormal ringing. We determined the quasinormal frequencies by
two different approaches, 6th order WKB and time domain integration
with Pronny fitting of the numerical data, obtaining by both methods very close numerical results. 

At very late times, the evolution of fermion perturbations in four dimensional stringy black holes shows a power law falloff proportional to $t^{-(2\ell+3)}$, as functions of the multipole number. This behavior is identical to that appeared in the late time evolution
of Dirac fields in other four dimensional black holes.

There are extensions of this work that would be considered in the future. In the first place it would be interesting to study of Dirac perturbations in higher dimensional stringy black holes.
Another interesting problem is related with the analytical investigation of the late-time behavior of massless Dirac perturbations outside stringy black holes, to obtain a result in correspondence with our numerical estimate for the decay factor. 
We also find interesting the consideration of the influence of vacuum polarization effects in the quasinormal spectrum for semiclassical stringy black holes. The solutions to some of the above
problems will be presented in future reports.

%%%%%%%%%%%%%%%%%%%%%%%%%%%%%%%
\section*{Acknowledgments}

This work has been supported by FAPESP (\emph{Funda\c c\~ao de Amparo \`a Pesquisa do Estado de S\~ao Paulo}) and CNPQ (\emph{Conselho Nacional de Desenvolvimento Cient\' ifico e Tecnol\'ogico}), Brazil.
  We are grateful to Professor Elcio Abdalla and Dr. Alexander Zhidenko for the useful suggestions and J. Basso Marques for technical support.
%%%%%%%%%%%%%%%%%%%%%%%%%%%%%%

\end{document}